\documentclass[aps,pra,showpacs,floatfix,onecolumn,superscriptaddress]{revtex4-2}

\usepackage{bm}
\usepackage{amsmath,amssymb,amsthm,times,graphics,graphicx,bm,xcolor}
\usepackage[normalem]{ulem}
\usepackage[colorlinks=true,linkcolor=blue,urlcolor=blue,citecolor=blue,pdfusetitle]{hyperref}
\usepackage{braket}

\newcommand{\SP}[1]{{\color{black}{#1}}} 
\newcommand{\TG}[1]{{\color{black}{#1}}}

\begin{document}

\title{Mode structure reconstruction by detected and undetected light}

\author{L. T. Knoll}
\affiliation{INRIM, Strada delle Cacce 91, I-10135 Torino, Italy}
\affiliation{DEILAP-UNIDEF, CITEDEF-CONICET, J.B. de La Salle 4397, 1603 Villa Martelli, Buenos Aires, Argentina}
\author{G. Petrini}
\affiliation{INRIM, Strada delle Cacce 91, I-10135 Torino, Italy}
\affiliation{Physics Department – University of Torino, Via Pietro Giuria 1, I-10126 Torino, Italy}
\author{F. Piacentini*}
\affiliation{INRIM, Strada delle Cacce 91, I-10135 Torino, Italy}
\author{P. Traina}
\affiliation{INRIM, Strada delle Cacce 91, I-10135 Torino, Italy}
\author{S. V. Polyakov}
\affiliation{National Institute of Standards and Technology, 100 Bureau Drive, Gaithersburg, Maryland 20899, USA}
\author{E. Moreva}
\affiliation{INRIM, Strada delle Cacce 91, I-10135 Torino, Italy}
\author{I. P. Degiovanni}
\affiliation{INRIM, Strada delle Cacce 91, I-10135 Torino, Italy}
\affiliation{INFN, sezione di Torino, via P. Giuria 1, 10125 Torino, Italy}
\author{M. Genovese}
\affiliation{INRIM, Strada delle Cacce 91, I-10135 Torino, Italy}
\affiliation{INFN, sezione di Torino, via P. Giuria 1, 10125 Torino, Italy}

\date{\today}

\begin{abstract}

We introduce a novel technique for the reconstruction of multimode optical fields, based on simultaneously exploiting both the generalized Glauber's $K^{th}$-order correlation function $g^{(K)}$ and a recently proposed anti-correlation function (dubbed $\theta^{(K)}$) which is resilient to Poissonian noise.
\TG{We experimentally demonstrate that this method yields mode reconstructions with higher fidelity with respect to those obtained with reconstruction methods based only on $g^{(K)}$'s, even requiring less ``a priori'' information.}
The reliability and versatility of our technique make it suitable for a widespread use in real applications of optical quantum measurement, from quantum information to quantum metrology, especially when one needs to characterize ensembles of single-photon emitters in the presence of background noise (due, for example, to residual excitation laser, stray light, or unwanted fluorescence).
	
\end{abstract}
\maketitle

\section{Introduction}
Recent years have seen an impressive advancement of quantum technology in the optical domain \cite{comm1,geno,genoAVS,sciarreview,coronachan} and single-photon metrology \cite{mig} up to a point where the associated techniques are not anymore restricted to scientific labs, but are starting to effectively proliferate to the industry \cite{comp, thierry} and the world-wide market \cite{qnami}, ultimately approaching everyday's life. This much-awaited ``second quantum revolution'' \cite{roadmap} paves the way for increasingly complex schemes to exploit the advantages of quantum effects for practical applications in practical scenarios such as quantum computation \cite{comp,comp2,comp3,comp4,comp5,comp6,comp7}, quantum communication \cite{pan,comm1,comm2}, quantum-enhanced measurement \cite{aligo,enh1,enh2,enh3,enh4,enh5}, 
quantum imaging and sensing \cite{ivano,imag1,barry,sens1,sens2,sens3}, and quantum testing \cite{test1,test2}.
As a consequence, it is of the utmost importance to develop simple methods \cite{m1,m4,m6,m7,m8,m9,m10} 
to characterize optical states that are significantly more complex than that of the proof-of-principle single isolated quantum systems (with possible addition of a small amount of background).
Composite and application-driven quantum systems require an appropriate characterization.
Such systems are significantly affected by inevitable noise and decoherence effects occurring when the system is moved from a controlled lab-like environment to a real-world one for a practical application.
From a theoretical point of view, devising reliable and robust nonclassicality criteria for such quantum systems is a topic of high interest \cite{rigovacca,sperling1,perina}. 
For instance, the characterization of ensembles of single-photon sources (SPSs) \cite{chun,sergey} 
in the presence of strong noise baths is considered.
The most widespread techniques for the characterization of quantum optical states are based on the measurement of second order Glauber's autocorrelation function, defined as
\begin{equation}
g^{(2)}(\tau)=\frac{\langle E^{(-)}(t)E^{(-)}(t+\tau)E^{(+)}(t+\tau)E^{(+)}(t)\rangle}{\langle E^{(-)}(t)E^{(+)}(t)\rangle^2}\,,
\end{equation}
and in particular its $g^{(2)}$(0) value.\\
This parameter is typically used to intuitively assess the nonclassicality of optical sources, since its value is below one for sub-poissonian non-classical light, equal to one for a Poissonian (laser) source, and above one for other classical states. 
In particular, $g^{(2)}(0)$ vanishes for a SPS, being exactly 0 in the ideal case \cite{g2}.
In the low-photon-flux regime, \TG{i.e. when $P(n+1) \ll P(n) \ll 1$ (being $P(n)$ the probability of observing $n$ photons in our detector)}, this parameter is equivalent to Grangier's parameter $\alpha$ \cite{grang}, defined as the ratio between the photon coincidence probability and the product of the single photon detection probabilities at the output of a Hanbury-Brown $\&$ Twiss interferometer (HBTI), which is the typical device used to measure $g^{(2)}$ experimentally.
This parameter ($g^{(2)}$ or $\alpha$, with no distinction in the following treatment) can immediately be extended to any order $K$ by defining $g^{(K)}$ as the ratio of the probability of a $K$-fold coincidence divided by the product of $K$ single click probabilities of $K$ non-photon-number-resolving (non-PNR) detectors attached to the output ports of a generalized multiport HBTI. Operationally, such an HBTI can be comprised of cascaded two-ports beam splitters \cite{spad,spad2,spad3,spad7,spad8,spad9}. 
One of the main advantages of this parameter is that its value does not depend on the splitting ratio among the HBTI arms, on the overall losses and on the detection efficiency of the detectors comprising the HBTI.

The experimental measurement of $g^{(K)}$ has proven to be a useful resource in quantum optics for several applications ranging from SPS characterization, quantum super-resolved imaging \cite{sres,sres2} and reconstruction of modal structure of composite optical fields \cite{gold}.
In this latter instance, it has been demonstrated how to identify, by simultaneously sampling multiple-order $g^{(K)}$'s (in the specific case, $K=2,3,4$), the underlying mode structure of complex multimode fields such as the superposition of a SPS emission with thermal fields, or a multi-thermal field with a Poissonian field, a task that cannot be achieved by only measuring $g^{(2)}$.
This useful technique presents some limitations, emerging for instance when the fields to be reconstructed are composed by one or more distinct SPSs in presence of Poissonian or both thermal and Poissonian background noise, a situation of interest, e.g., when identifying single-photon emission from color centers in diamond.
Furthermore, some ``a priori'' knowledge on the state to be reconstructed (e.g., the number and types of modes composing it) is needed to achieve reliable results.
In some cases, particularly with true PNR detectors, the use of the set of probabilities $\{p_K\}$ to detect photon states with up to $K$ photons is useful \cite{buren, burenSITE}, but because those values depend on loss including the detection efficiency and detector saturation, advanced characterization of detectors may be required.

Lately, a new criterion for assessing optical sources nonclassicality, mainly focused on clusters of single-photon emitters, has been proposed \cite{filip} and successfully implemented to test SPSs based on emitters such as color centers \cite{prb}, trapped ions \cite{filip2} and colloidal CdSe/CdS dot-in-rods\cite{Andrea}.
This criterion is based on the measurement of a parameter, $\theta^{(K)}$, defined as:
\begin{equation}\label{eq:deftheta}
\theta^{(K)}(0)=\frac{Q(0^{\otimes K})}{(Q(0))^K},
\end{equation}
where $Q(0)$ and $Q(0^{\otimes K})$ are, respectively, the probability of no-photon detection at the end of one arm and in $K$ HBTI arms simultaneously.
The parameter $\theta^{(K)}$ has two main interesting properties: first, the $\theta^{(K)}$ value
is not affected by the presence of Poissonian light, so that it can be extremely valuable in the characterization of
photoluminescent emitters \cite{CCiD1,CCiD3,CCiD5,CCiD6,CCiD7,CCiD10,CCiD12,masha}, 
since such a parameter would be insensitive to residual back-reflected excitation laser light.
Second, in contrast with $g^{(K)}$, when characterizing clusters of SPSs the $\theta^{(K)}$ value decresases as the number of emitters in the ensemble increases.
This property is of special interest when characterizing large ensembles, since $g^{(K)}\rightarrow1$ for $K\rightarrow\infty$.
As a drawback, $\theta^{(K)}$, contrarily to $g^{(K)}$, strongly depends on the experimental apparatus, i.e. the BSs splitting ratio, the optical transmission of the HBTI and detection efficiencies of the detectors involved.

The aim of this work is presenting an innovative method for the reconstruction of optical states exploiting both the $g^{(K)}$ and $\theta^{(K)}$ parameters simultaneously. This method outperforms the mode reconstruction technique exploiting only $g^{(K)}$'s \cite{gold} in terms of robustness as well as versatility, and is particularly advantageous for  measurements in HBTIs arrangements with non-PNR detectors \SP{(because $\theta^{(K)}$'s are insensitive to Poissonian fields, a reconstruction method that only uses $\theta^{(K)}$'s cannot generally grant reliable results)}.
We will show how the combined approach can provide a reliable quantitative evidence of single-photon emission even in presence of strong classical (thermal and/or Poissonian) light.

\section{Theoretical model}
The physical system considered here (see Fig. \ref{fig:exp_setup}) is the emission of multimode light from one or many different optical sources observed by $N=4$ non-PNR detectors in a tree configuration.
For simplicity, we assume that photons are split to $N$ branches of a detector tree with equal probability $1/N$, and that each detector has identical system efficiency (including transmission losses and detection efficiency) $\eta$.
This assumption does not qualitatively change the results.
Let us define 
the characteristic function 
for a discrete probability function $p_n$ (with $\sum_{n=0}^{+\infty}p_n=1$)
\begin{equation}\label{eq:funzioni}
\Gamma(z)=\sum_{n=0}^{+\infty} p_n z^n
\end{equation}
Accounting for the efficiency $\eta$, the characteristic function in Eq. \eqref{eq:funzioni} becomes
\begin{equation}\label{eq:funzioni_eta}
\Gamma(z)=\sum_{n=0}^{+\infty}[1-\eta(1-z)]^n p_n
\end{equation}
The characteristic function $\Gamma(z)$ has the following properties:
%
\begin{equation}
    \begin{array}{r@{}l}
\label{eq:propgamma}
\frac{d}{dz} \Gamma (z)\biggr|_{z=1} &{} =\mathcal{E}\{n\}, \\
\frac{d^2}{dz^2} \Gamma (z)\biggr|_{z=1} &{} =\mathcal{E}\{n(n-1)\}, \\
\vdots\\
\frac{d^K}{dz^K} \Gamma (z)\biggr|_{z=1} &{} =\mathcal{E}\big\{\frac{n!}{(n-K)!}\big\}, \\
\end{array}
\end{equation}
where $\mathcal{E}\{x\}$ represents the expectation value of the variable $x$.
It is straightforward to show from Eqs. (\ref{eq:propgamma}) that, for a single optical mode, the generic $g^{(K)}(0)$ function can be expressed as:
\begin{equation}
g^{(K)}(0)=\frac{\frac{d^K}{dz^K} \Gamma(z)}{\bigl(\frac{d}{dz} \Gamma(z)\bigr)^K}\biggr|_{z=1}.
\label{eq:gK}
\end{equation}
Let us now suppose that we have a combination of several optical sources at once, with different statistical distributions.
For instance, for $M$ single-photon emitters with photon emission probability $p$, one thermal source and one Poissonian source, the total photon-number probability distribution reads:
%
%
\begin{equation}\label{pntot}
p_n^{TOT}=\sum_{k,l,m} \delta_{n,(k+l+m)}P_k^{bin}(p,M)P_l^{th}(\nu)P_m^{poi}(\mu),
\end{equation}
where it has been assumed that the three generated fields have, respectively, binomial $\big(P_k^{bin}(p,M)=\binom{M}{k}p^k(1-p)^{M-k}\big)$, thermal $\big(P_l^{th}(\nu)=\frac{\nu^l}{(1+\nu)^{l+1}}\big)$ and Poissonian $\big(P_m^{poi}(\mu)=\frac{\mu^m e^{-\mu}}{m!}\big)$ photon number distributions.
\SP{In general, the total photon-number probability (c.f. Eq. \eqref{pntot}) distribution can be obtained for any number and type of emitters in a similar fashion.}
For a multimode field in Eq. \eqref{pntot}, the statistical distribution of each mode is given by a characteristic function, and the composite characteristic function $\Gamma^{TOT}(z)$ can be written as
\begin{equation}
\Gamma^{TOT}(z) = \Gamma^{th}(z) \Gamma^{poi}(z)\Gamma^{bin}(z),
\label{eq:factor}
\end{equation}
where $\Gamma^{th}(z)$, $\Gamma^{poi}(z)$ and $\Gamma^{bin}(z)$ are, respectively, the characteristic functions related to the thermal source, the coherent one and the SPS ensemble (see eq. \eqref{eq:funzioni_eta}).
Thus, the $g^{(K)}(0)$ values can be calculated from the  characteristic function $\Gamma^{TOT}(z)$ from Eq. \eqref{eq:gK} and Eq. \eqref{eq:factor}.
In this paper we reconstruct unknown multimode optical fields comprised of $S$ modes.
Instead of requiring a priori knowledge on the photon number statistics of each of the $S$ modes, we identify all possible $S$-mode combinations with arbitrary statistics:
\begin{equation}
\label{eq:gammaN}
 \Gamma^{TOT}(z) = \delta_{S,(j_{poi}+j_{bin}+j_{th})} \prod_{j=1}^{j_{poi}}\Gamma_j^{poi}(z)\prod_{j=1}^{j_{th}}\Gamma_j^{th}(z)\prod_{j=1}^{j_{bin}}\Gamma_j^{bin}(z)
\end{equation}
(since the sum of multiple Poisson distributions is also a Poisson distribution, it is enough to consider only one Poissonian mode, i.e. $j_{poi}\leq1$).
To find the right reconstruction, we compare fit quality for all models and chose the best one, for details see the Results and Methods sections.
Let us now investigate the expression of the $\theta^{(K)}(0)$ function, defined in Eq. (\ref{eq:deftheta}), for this multimode field.
The no-click probability of the detector at the end of the $i^{th}$ branch of a $K$-branch detector tree with $n$ impinging photons can be calculated as the convolution of the probability of having $k_i$ out of $n$ photons in the $i^{th}$ branch (governed by binomial distribution) and the probability of observing zero out of $k_i$ incoming photons in the same branch ($\pi_i=(1-\eta)^{k_i}$, where $\eta$ is the detection efficiency of the detector):
\begin{equation}\label{eq:q0}
Q_i(0|n) = \sum\limits_{k_i=0}^{n} \frac{n!}{n!(n-k_i)!}\biggl(\frac{1-\eta}{K}\biggr)^{k_i} \biggl( 1-\frac{1}{K}\biggr)^{n-k_i} =\biggl( 1-\frac{\eta}{K}\biggr)^n.
\end{equation}
Analogously, the probability of detecting zero out of $n$ photons simultaneously in $K\leq N$ branches is the probability of a particular permutation of $n$ photons over $K$ branches of the detector-tree (governed by the multinomial distribution) multiplied by the joint probability of detecting zero photons in each branch ($\prod_i^K \pi_i$) considering all the possible photon distributions in the $K$ branches, i.e. all possible $\{k_i\}$ sets fulfilling the condition $\sum_{i=1}^{K} k_i=n$:
\begin{equation}\label{eq:q0n}
Q(0^{\otimes K}|n) =  \sum\limits_{k_1,\dots,k_K \atop \left[\sum_{i=1}^{K}k_i=n\right]}\frac{n!}{k_1! \dots k_K!}\prod\limits_{i=1}^K \biggl( \frac{1-\eta}{K} \biggr)^{k_i}=\biggl(\sum_{i=1}^{K}\frac{1-\eta}{K}\biggr)^n=\left(1-\eta\right)^n\,.
\end{equation}
In order to calculate the terms in Eq. (\ref{eq:deftheta}), the conditional probabilities in Eqs. (\ref{eq:q0}) and  (\ref{eq:q0n}) must be averaged over the statistical distribution.
In this case it can be shown, with a procedure analogous to the one of Eq. (\ref{eq:factor}), that $Q^{TOT}(0)$ and $Q^{TOT}(0^{\otimes K})$ can be factorized.
\SP{Here we show factorization using the same example of $m$ single photon sources, one thermal and one Poissonian mode (c.f. Eq. \eqref{pntot}).}
\begin{equation}
Q^{TOT}(0) = \sum_{n=0}^{\infty}(1-\frac{\eta}{K})^n p_n^{TOT} = Q^{th}(0) Q^{poi}(0)Q^{bin}(0)
\end{equation}
and
\begin{equation}
Q^{TOT}(0^{\otimes K}) = \sum_{n=0}^{\infty}(1-\eta)^n p_n^{TOT} = Q^{th}(0^{\otimes K}) Q^{poi}(0^{\otimes K})Q^{bin}(0^{\otimes K}).
\end{equation}
Thus, the $\theta^{(K)}(0)$ function can be calculated as:
\begin{equation}
\theta^{(K)}(0)=\frac{Q^{th}(0^{\otimes K}) Q^{poi}(0^{\otimes K})Q^{bin}(0^{\otimes K})}{(Q^{th}(0) Q^{poi}(0)Q^{bin}(0))^K}=\frac{Q^{th}(0^{\otimes K}) Q^{bin}(0^{\otimes K})}{(Q^{th}(0)Q^{bin}(0))^K},
\label{eq:thetaN}
\end{equation}
where we used the property $Q^{poi}(0^{\otimes K})=[Q^{poi}(0)]^K$, making $\theta^{(K)}$ insensitive to Poissonian light and, as a consequence, resilient to a Poissonian noise.
Again, this result can be extended to an arbitrary number of sources of each type by simply including appropriate multipliers to the above factorized expression, in a similar manner to Eq. \eqref{eq:gammaN}.


\section{Results}
\begin{figure}[htp!]
	\centering
	\includegraphics[width=0.7\columnwidth]{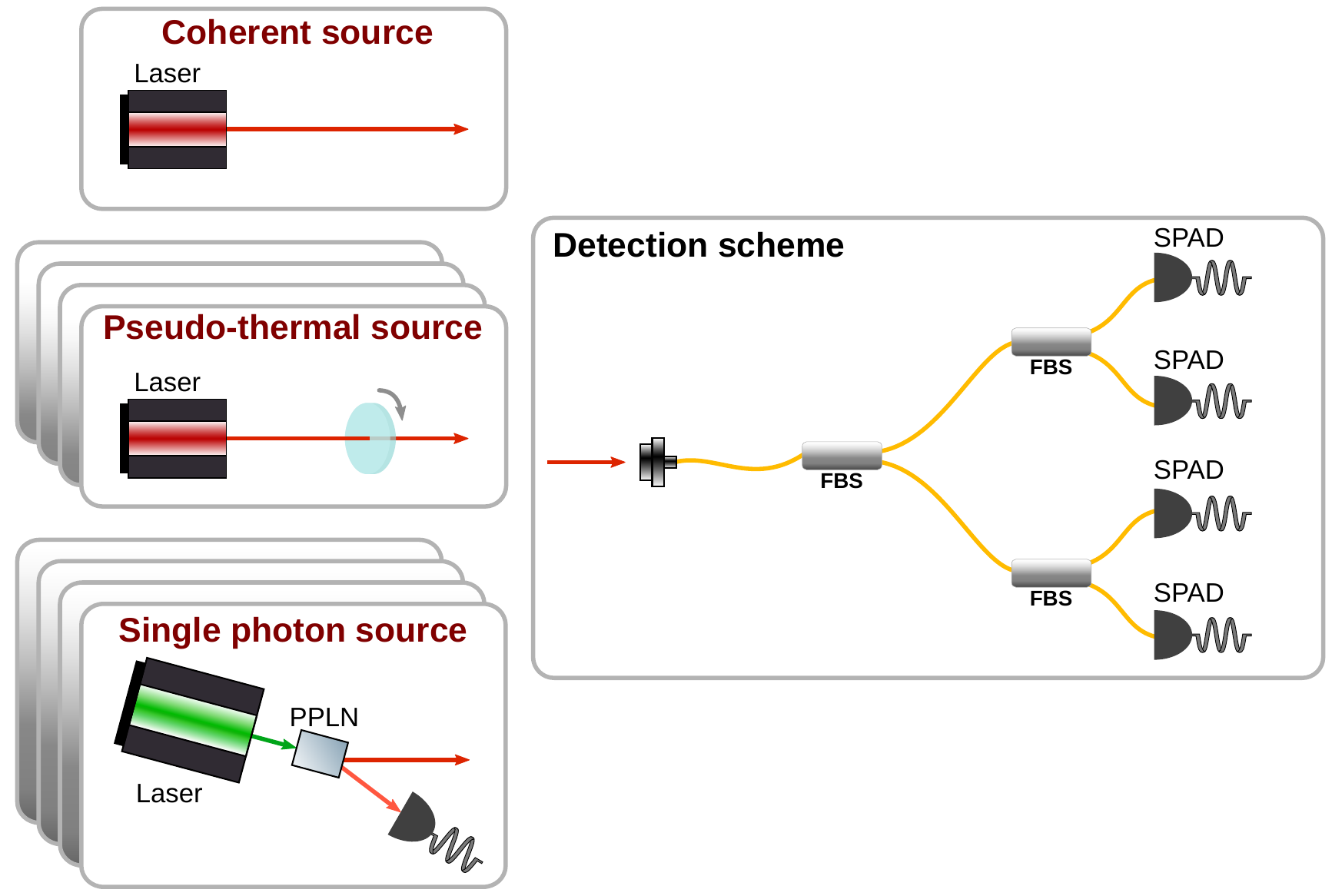}
	\caption{Experimental setup. Faint states of light under study are a classical or non-classical multimode fields.
The non-classical fields correspond to the emission of $M\leq N$ single-photon sources, with a strong thermal and/or Poissonian noise added.
The classical fields, instead, are arbitrary compositions of multiple thermal modes and a Poissonian mode.
On the left, three types of sources generating faint light at $1.55$ $\mu$m are shown: a coherent (Poissonian) mode, produced by attenuating a pulsed laser;  pseudo-thermal mode(s), generated by the pulsed laser sent through a rotating ground glass disk; single-photon mode(s) are emitted by a heralded single-photon source based on SPDC in a PPLN crystal.
On the right, a pictorial scheme of our detector tree, which consists of a cascade of three 50:50 fiber beam splitters (FBSs) in a tree configuration connected to four InGaAs/InP SPADs is shown.}
	\label{fig:exp_setup}
\end{figure}

\begin{figure}[htp!]
	\centering
	\includegraphics[width=0.7\columnwidth]{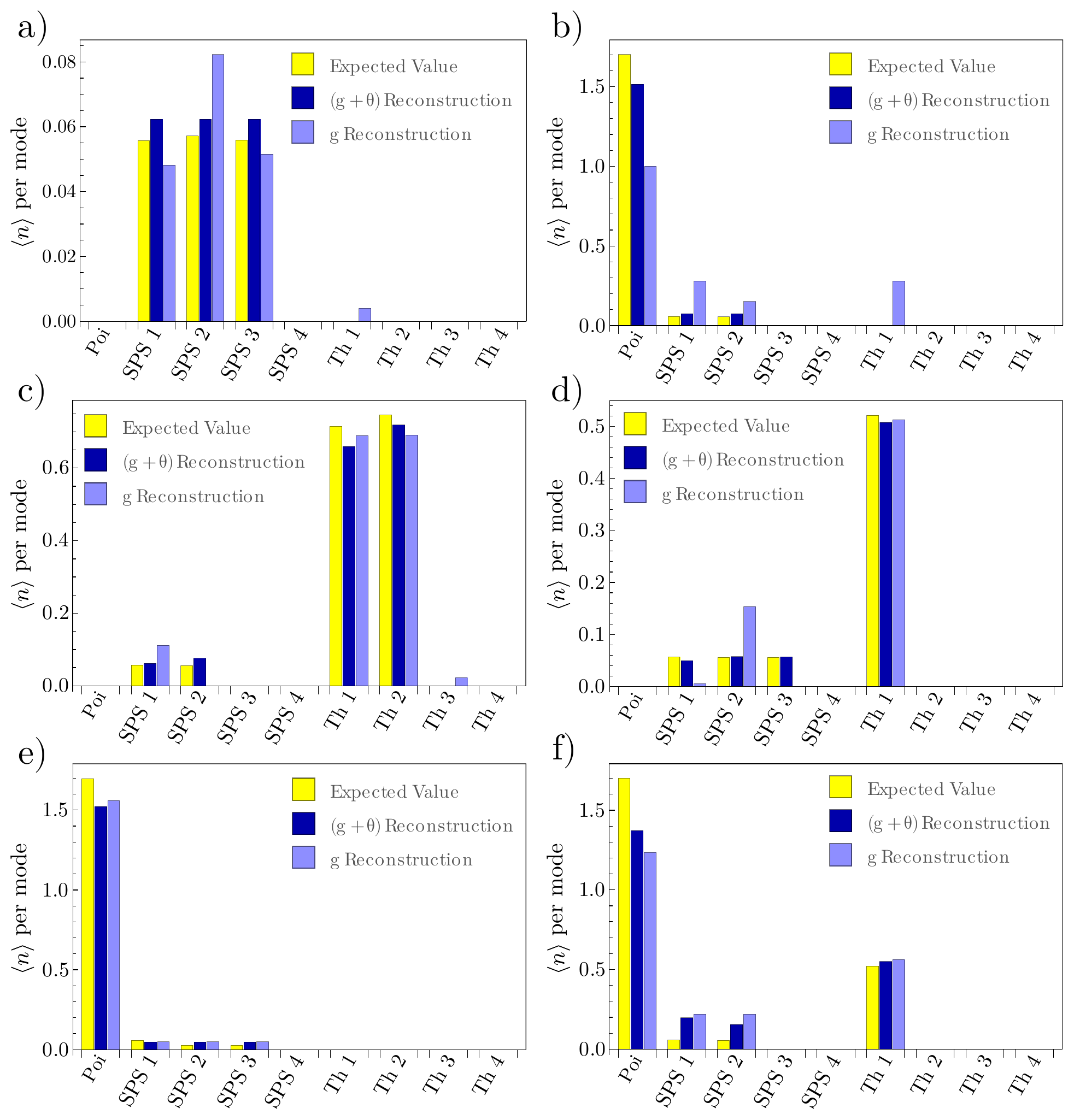}
	\caption{Mode reconstruction results. Expected and reconstructed mean photon number per mode for light fields generated by: a) three single-photon emitters; b) two single-photon emitters in the presence of Poissonian light; c) two single-photon emitters together with two thermal fields; d) three single-photon emitters in the presence of a thermal field; e) three single-photon emitters in the presence of a Poissonian field; f) two single-photon emitters with both a Poissonian and a thermal mode.
Yellow bars correspond to the mean photon numbers per mode present in light field under measurement, whilst dark and light blue bars represent, respectively, the ones obtained with the $g^{(K)}+\theta^{(K)}$ and $g^{(K)}$-only reconstruction techniques. Poi: Poissonian mode. SPS: single photon state. Th: thermal mode.}
	\label{fig:6plots}
\end{figure}

We test our technique by applying it to several different multi-mode optical fields detected by our detector tree.
The detector tree is composed of three 50:50 fiber beam splitters (FBSs) connected to $N=4$ InGaAs/InP single-photon avalanche diodes (SPADs) in a tree configuration, allowing us to discriminate up to four incoming photons.\\
We generate a multimode optical field combining, in our experimental setup (Fig. \ref{fig:exp_setup}), up to four optical modes (because of the $N=4$ constraint on the detection side).
The multimode field is produced by three different source types: a Poissonian (coherent) source, thermal source and single-photon source (see Methods for details).
%
The reconstruction of the $S$-mode field is achieved by a minimization algorithm based on a least square difference between the theoretical $g^{(K)}$ and $\theta^{(K)}$ (labeled ``rec'') and the ones obtained in the experiment (labeled ``exp'').
Specifically, the function to be minimized is
\begin{equation}\label{eq:Fmin}
LS=\sum_{K=2}^{4}\lambda_{g}(K)\left(g^{(K)}_{\mathrm{rec}}(0)-g^{(K)}_{\mathrm{exp}}(0) \right)^2 + \lambda_{\theta}\sum_{K=2}^{4}\left(\theta^{(K)}_{\mathrm{rec}}(0)-\theta^{(K)}_{\mathrm{exp}}(0) \right)^2\,,
\end{equation}
where $\lambda_{\theta}$ and $\lambda_g(K)$ are Lagrange multipliers, $g^{(K)}_{\mathrm{rec}}(0)$ is given by Eq. \eqref{eq:gK} and $\theta^{(K)}_{\mathrm{rec}}(0)$ by Eq. \eqref{eq:thetaN}.
The photon number resolution of our detection system is limited to $N=4$, therefore we can only reconstruct a maximum of $S=4$ arbitrary modes.
In our reconstruction algorithm we assume that these four modes are unknown.
So, with each experimental data set, we perform mode reconstructions for all possible four-mode combinations of one Poissonian, four thermal and four single-photon modes.
We then compare the minimized LS values and choose the mode combination and the set of reconstructed average energies per mode that result in the lowest LS value (details in Methods).
In this way, our algorithm truly identifies the multimode light field with unknown modes, and not merely matches the previously-known modes with appropriate mean photon numbers.
To test the robustness and reliability of our method, in our experiment we perform a series of acquisitions in several regimes, combining different modes and comparing the results of our mode-reconstruction technique (exploiting both $\theta^{(K)}$ and $g^{(K)}$ parameters) with the ones obtained using only the $g^{(K)}$ (adding in both cases a further constraint on the overall no-click probability $Q(0^{\otimes N})$, to define the average number of photons of the light field) as in \cite{gold}.
In particular, we especially focus on cases in which the multi-mode optical field under test features one or more single-photon modes, heavily polluted by classical (thermal and/or Poissonian) light, giving an overall $g^{(2)}(0)\geq1$.
%
The obtained results are summarized in Table \ref{tab:Fidelity}, comparing the fidelity achieved by both reconstruction methods, defined as the distance $F_x=(2|\vec{m}_\mathrm{e}\cdot\vec{m}_x|$)/$(|\vec{m}_\mathrm{e}|^2+|\vec{m}_x|^2)$, where $\vec{m}_\mathrm{e}$ is the set of expected mean photon numbers in each mode and $\vec{m}_x$ is the one reconstructed with the ``$x$'' method ($x=g+\theta$ labels the one exploiting both $g^{(K)}$ and $\theta^{(K)}$ functions, whilst $x=g$ indicates the one based solely on $g^{(K)}$).
The expected mean photon number set $\vec{m}_\mathrm{e}$ is obtained by separately measuring the mean photon number per pulse of each mode composing the optical field to be reconstructed.
Table \ref{tab:Fidelity} shows the number of modes present in the multimode light field under examination ($S^{e}$), as well as the number of modes identified by both reconstruction methods, respectively labeled $S^{\mathrm{rec}}_{g+\theta}$ and $S^{\mathrm{rec}}_{g}$ (for both of them, the number of correctly recognized modes' types is indicated in parentheses).
For each case studied, the value of $g^{(2)}_{\mathrm{exp}}(0)$, the observable that is typically used for discriminating between classical and non-classical states, is also reported in the last column of Table \ref{tab:Fidelity}.
\begin{table}[htbp]
\begin{center}
\begin{tabular}{||c||c||c||c|c||c|c||c||}
\hline
 & &  &   \multicolumn{2}{c||}{Method $g+\theta$} &  \multicolumn{2}{c||}{Method $g$} &    \\
$\;$ Case $\;$ & $\;$(a)$\;$ $\;S^{\mathrm{e}}\;$ & $\;$(b)$\;\;$ Mode configuration $\,\,\,$ & $\;$(c)$\;\;$ $F_{g+\theta}\;\;$ & $\;$(d)$\;$ $\;S^{\mathrm{rec}}_{g+\theta}\;$  &  \;(e)$\;\;$ $F_g\;\;$ & $\;$(f)$\;\;$ $\;S^{\mathrm{rec}}_g\;$  & $\;$(g)$\;\;$ $g^{(2)}_{\mathrm{exp}}(0)$   \\
\hline
\hline
I & 4 & 1 SPS, 2 Th, 1 Poi	& 0.9597 & 4 (4) & 0.9337 & 4 (4) & $\;\;1.137\pm0.002\;\;$\\
II & 4 & 1 SPS, 3 Th		& 0.9518 & 4 (4) & 0.9480 & 4 (4) & $1.332\pm0.002$ \\
III & 4 & $\;\;$2 SPS, 1 Th, 1 Poi $\star\;\;$	& 0.9745 & 4 (4) & 0.9469 & 4 (4) & $1.044\pm0.003$\\
IV & 4 & 2 SPS, 2 Th $\star$	& 0.9979 & 4 (4) & 0.9949 & 4 (3) & $1.411\pm0.005$\\
V & 4 & 3 SPS, 1 Poi $\star$	& 0.9941 & 4 (4) & 0.9963 & 4 (4) & $0.998\pm0.003$\\
VI & 4 & 3 SPS, 1 Th $\star$	& 0.9996  & 4 (4) & 0.9729 & 3 (3) & $1.532\pm0.012$\\
VII & 4 & 3 Th, 1 Poi		& 0.9819 & 4 (4) & 0.7325 & 4 (3) & $1.103\pm0.001$\\
VIII & 4 & 4 Th			& 0.9547 & 4 (4) & 0.8481 & 4 (3) & $1.245\pm0.001$\\
IX & 3 & 1 SPS, 1 Th, 1 Poi	& 0.9885 & 3 (3) & 0.9755 & 4 (3) & $1.072\pm0.002$\\
X & 3 & 1 SPS, 2 Th		& 0.9934 & 3 (3) & 0.9390 & 3 (3) & $1.478\pm0.003$\\
XI & 3 & 2 SPS, 1 Poi $\star$	& 0.9931 & 3 (3) & 0.8463 & 4 (3) & $0.996\pm0.004$\\
XII & 3 & 2 SPS, 1 Th		& 0.9972  & 3 (3) & 0.8325 & 3 (2) & $1.732\pm0.011$\\
XIII & 3 & 2 Th, 1 Poi		& 0.9749 & 3 (3) & 0.9749 & 4 (3) & $1.135\pm0.001$\\
XIV & 3 & 3 SPS $\star$		& 0.9947 & 3 (3) & 0.9660 & 4 (3) & $0.64\pm0.03$\\
XV & 3 & 3 Th			& 0.9509 & 3 (3) & 0.9490 & 3 (3) & $1.349\pm0.001$\\
\hline
\end{tabular}
\caption{Performance comparison between the $g+\theta$ and the $g$ methods.
Columns (a) and (b) show, respectively, the number $S^{\mathrm{e}}$ of modes and mode types of the multimode light field under measurement and subsequent reconstruction.
Column (c) shows the fidelity $F_{g+\theta}$ between the expected multimode optical field and the one reconstructed exploiting both $g^{(K)}$ and $\theta^{(K)}$, while column (d) indicates $S^{\mathrm{rec}}_{g+\theta}$, i.e. the number of optical mode types identified (correctly identified) by this technique.
Columns (e) and (f) are same as columns (c) and (d), respectively, but for the reconstruction method that uses $g^{(K)}$'s only.
Finally, column (g) shows the $g_{\mathrm{exp}}^{(2)}(0)$ value experimentally measured for each mode configuration.
Stars indicate reconstructions depicted in Fig. \ref{fig:6plots}, and graphically compared to theoretically-expected values \TG{(the reconstruction plots pertaining to all configurations, together with the expected counterparts, are reported in the Supplemental Material)}.
SPS: single photon state; Th: thermal mode; Poi: Poissonian mode.}
\label{tab:Fidelity}
\end{center}
\end{table}
These results demonstrate that combining $g^{(K)}$ and $\theta^{(K)}$ method manages to faithfully reconstruct the modal structure of the multimode light field characterized by our detector tree, always obtaining large fidelities (above 0.95) and identifying the correct number and type of optical modes for all the cases investigated.
This gives the experimental proof of both the reliability and robustness of our method, that clearly outperforms the one relying solely on the $g^{(K)}$ \cite{gold} in all the cases (except for case V where fidelities are comparable).
The latter, in fact, not only achieves comparatively lower fidelities (occasionally going below 0.9, indicating poor reconstruction), but in half of the cases it does not correctly identify the number and types of optical modes comprising the multimode field under test, as it is evident from column (f) of Table \ref{tab:Fidelity}.\\
The expected and reconstructed modal structures for the multi-mode fields marked with a star in Table \ref{tab:Fidelity}, column (b), are shown in Fig. \ref{fig:6plots}.
We show selected cases in which different single photon emitters are combined together (Fig. \ref{fig:6plots}a) or mixed with strong Poissonian and/or thermal sources (plots \ref{fig:6plots}b-\ref{fig:6plots}f; see Supplemental Material for all the other results of mode reconstruction).
Each plot compares the mean-photon number of every mode present in the light field (yellow bars) with the reconstructed one obtained with our novel technique (dark blue bars) and with the reconstruction method exploiting only the $g^{(K)}$'s (light blue bars).
In particular, Fig. \ref{fig:6plots} shows the following cases: a) three single-photon emitters; b) two single-photon emitters in presence of heavy Poissonian noise; c) two single-photon emitters in presence of two thermal sources; d) three single-photon emitters in presence of thermal noise; e) three single-photon emitters in presence of heavy Poissonian noise; f) two single-photon emitters mixed with both a Poissonian and a thermal mode.
Even though the Poissonian and thermal mode intensities are, respectively, about 30 and 10 times higher than that of each single-photon emitter, our technique correctly recognizes and reconstructs the type and number of light modes composing our optical field, identifying non-classical single-photon emission even in ostensibly classical optical fields, i.e. with $g^{(2)}_{\mathrm{exp}}(0)\geq1$, and not finding single-photon emission in multimode fields with no single-photon mode.
\TG{The residual mismatch between expected and reconstructed mode structures is reasonably due to imperfections in the detection apparatus, such as, e.g., dark counts, discrepancies in the detector tree branches and in their efficiency estimation, and the higher statistical uncertainty associated to $g^{(K)}$ and $\theta^{(K)}$ for high $K$.}

\section{Discussion}
Overall, our technique exploiting both $g^{(K)}$ and $\theta^{(K)}$ enables a reliable reconstruction of the mode structure of very complex multimode fields, with simultaneous presence of Poissonian, thermal and/or single-photon emission, even in cases that are not successfully reconstructed with the method exploiting $g^{(K)}$ only.
This is particularly interesting, since it is a well known issue that, when sampling $g$-function only, it is practically impossible to distinguish the emission of a SPS in the presence of noise from the simultaneous emission of two distinct and differently coupled SPSs.
The studied cases demonstrate that the proposed technique is extremely efficient for characterizing SPSs in noisy environments, with practical applications to nonclassical emission from fluorescent targets.
The applications range from characterization of color centers in diamond \cite{CCiD1,CCiD3,CCiD5,CCiD6,CCiD7,CCiD10,CCiD12}, 
which can be affected (or even overtaken) by both Poissonian (residual excitation laser light) and thermal (stray light, unwanted fluorescence) noise contributions, to nonclassical imaging with fluorophores.
%
%
%
%
According to our results, the proposed technique for the mode reconstruction of optical fields, based on the combination of $g^{(K)}$ and $\theta^{(K)}$, not only outperforms the one illustrated in \cite{gold}, but it is also capable to reconstruct more complex mode structures that could not be processed with the legacy method, ultimately proving that supplying $\theta^{(K)}$ values to the mode reconstruction algorithm leads to superior performance.
Finally, this new technique does not rely on any ``a priori'' assumption on the number and type of modes constituting the optical field (except for the obvious constraint on the maximum number of modes allowed, due to the finite photon number resolution of the PNR detector used) not only is a clear evidence of its robustness, but also allows for its widespread application to several practical scenarios in quantum metrology and other quantum technologies.


\section*{Acknowledgments}
This work was supported by EMPIR projects 20FUN05 ``SEQUME'' and 20IND05 ``QADeT'' (these projects have received funding from the EMPIR programme co-financed by the Participating States and from the European Union's Horizon 2020 research and innovation programme), and by the European Commission's EU Horizon 2020 FET-OPEN project grant no. 828946 ``PATHOS''.
This work was also funded by the project QuaFuPhy (call ``Trapezio'' of Fondazione San Paolo).




\begin{thebibliography}{10}
\bibitem{g2}
Strangely enough, the value of  $g^{(2)}(0)$ is greater than zero for $n$-photons Fock states with $n\ge 1$, approximating 1 for large $n$, as in the case of large clusters of single photon emitters. This is to be considered an issue in the use of the $g^{(K)}$ parameters, since a Fock state with increasingly large $n$ does not approach a classical state. For this reason, this parameter alone is not optimal for the characterization of clusters of SPSs
\end{thebibliography}

\vspace{2cm}

\section*{Mode structure reconstruction by detected and undetected light: Supplemental Material}

\subsection*{Experimental setup}
In our experimental setup, shown in Fig. \ref{fig:exp_setup}, a pulsed telecom laser ($1.55$  $\mu$m) attenuated to the single-photon level generates a Poissonian mode.
Each pseudo-thermal mode is produced by making the same laser pass through a rotating ground glass disk.
Finally, heralded single-photon states at $1.55$ $\mu$m are obtained from a heralded single-photon source based on spontaneous parametric down-conversion (SPDC). A continuous wave (CW) laser (at $532$ nm) pumps a periodically-poled lithium niobate (PPLN) crystal, generating photon pairs at $810$ nm (idler) and $1.55$ $\mu$m (signal) \cite{sorgOE}. 
The idler photon is spectrally filtered and coupled to a single-mode fiber (SMF) connected to a Si-SPAD, heralding the presence of the corresponding signal photon. The generated state is close to a single-photon Fock state, with $g^{(2)}(0)<0.05$.
Once these modes are incoherently combined, the resulting multi-mode field is sent to our detector tree, allowing for a photon-number resolution up to $N=4$ photons. 
With our scheme, we can generate $S$-mode optical fields (with $S=1,...,N$) whose underlying mode structure can comprise up to $S$ thermal and/or single-photon modes, and up to one Poissonian mode, giving rise to $(2S+1)$ possible different modal configurations.
As stated above, in our particular case we consider a maximum of $N=4$ modes combined together.

\subsection*{Reconstruction algorithm}
In our least square minimization function (Eq. \eqref{eq:Fmin}), the $g^{(K)}_{\mathrm{exp}}$ are calculated as the ratio between the $K$-fold coincidence probability $Q_{(i_1,...,i_K)}(1)$ and the product of the single detection probabilities $Q_{i_1}(1),...Q_{i_K}(1)$ of the $K$ SPADs involved, averaged for all possible SPADs combinations.
The $\theta^{(K)}_{\mathrm{exp}}$ are evaluated from the overall no-click probability $Q_{(i_1,...,i_K)}(0)$ and the single branch no-click probabilities $Q_{i_j}(0)$ ($i_j=1,...,N$, with $j=1,...,K$) of the SPADs considered.
As stated above, while the $g^{(K)}$ by construction do not depend on the efficiency of detectors involved in their measurement, the same does not hold for the $\theta^{(K)}$ functions, which are intrinsically $\eta$-dependent.
For this reason, we took the efficiency unbalance between the branches comprising our detector tree in account by computing six different $\theta^{(2)}(0)$ values, four $\theta^{(3)}(0)$'s and one $\theta^{(4)}(0)$, each corresponding to a different combination of the detector-tree branches.\\
Furthermore, Lagrange multipliers are introduced in Eq. \eqref{eq:Fmin} for both $g^{(K)}$ and $\theta^{(K)}$ functions.
In particular, for each $g^{(K)}$ a different Lagrange multiplier $\lambda_g(K)$ is used according to the following rule:
\begin{equation}
 \lambda_g(K)=\left\{ \begin{array}{l l }
                      1/{K!} & \,\,\,\mathrm{if} \,\,\, g_{\mathrm{exp}}^{(2)}(0)>1\\
                      1 & \,\,\, \mathrm{otherwise.}  \\
                     \end{array}
                     \right.
\end{equation}
Whenever $g_{\mathrm{exp}}^{(2)}(0)\leq 1$, a unity Lagrange multiplier is applied to all Glauber functions ($\lambda_g(K)=1\,\,\,\forall\, K$).
Otherwise, we divide the corresponding square difference by the value $K!$, accounting for the factorial growth of $g^{(K)}(0)$ with $K$ for thermal modes.
In addition, the higher the order of an experimentally measured Glauber function, the higher is the associated uncertainty.
Thus, order-dependent Lagrange multipliers reduce the impact of uncertainties for large $K$'s.
\SP{The Lagrange multiplier $\lambda_\theta$ is found through a recursive algorithm.
First, $\lambda_\theta=1$ and the fit is obtained.
Then, $\lambda_\theta$ is adjusted and a new fit is performed such that the first term becomes equal to the second term in Eq. \eqref{eq:Fmin} through iterations.
Such an adjustment ensures that both $g^{(K)}$ and $\theta^{(k)}$ contributions to the cost function are similar.}\\
Finally, to increase the robustness and reliability of our reconstruction method, we use single-branch no-click probabilities $Q_{i}(0)$ as a constraint on the overall mean photon number of the reconstructed state.
Our minimization is carried on any four of nine unknown parameters, each characterizing a different source: $\mu$ is the mean photon number for the coherent mode, $\nu_1,\dots,\nu_4$ are the ones for the thermal modes and $p_1,\dots,p_4$ are the emission probabilities of the single-photon emitters for the single-photon modes.

\subsection*{Extended experimental results}
\begin{figure*}[htp!]
	\centering
	\includegraphics[width=17cm]{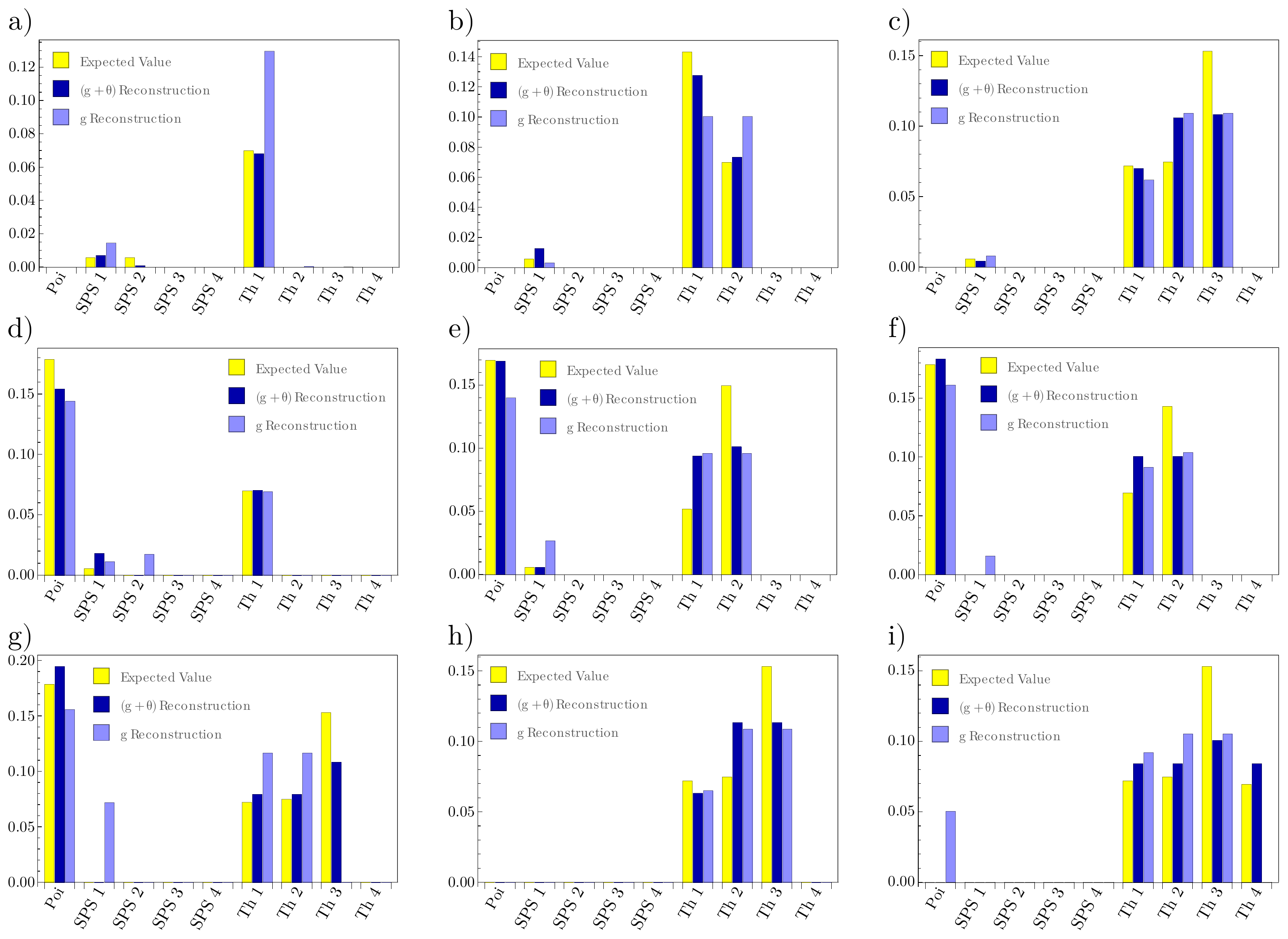}
	\caption{Reconstructed modes for a) two single-photon emitters in presence of a thermal source; b) one single-photon emitter and two thermal modes; c) one single-photon emitter and three thermal modes; d) one single-photon emitter, one Poissonian mode and one thermal mode; e) one single-photon emitter, one Poissonian mode and two thermal modes; f) two thermal modes in presence of a Poissonian source; g) three thermal modes in presence of a Poissonian source; h) three thermal modes and i) four thermal modes.
	Each bar corresponds to the mean-photon number for each mode present in our light field (yellow bars), the reconstructed one obtained with our technique (dark blue bars) and the one exploiting only the $g^{(K)}$'s (light blue bars), all in terms of the Poissonian (Poi), single-photon (SPS) and thermal (Th) components.}
	\label{fig:suppMat}
\end{figure*}

The plots for the reconstructed optical fields reported in Table \ref{tab:Fidelity} that where not shown in Fig. \ref{fig:6plots} are presented in Fig. \ref{fig:suppMat}.
The expected mean photon number for each configuration (yellow bars) is plotted along with the results obtained with our technique (dark blue bars) and with the one exploiting only the $g^{(K)}$'s (light blue bars), all in terms of the Poissonian, single-photon and thermal components.
Fig. \ref{fig:suppMat} a) shows two single-photon emitters in presence of a thermal source while b) one single-photon emitter and two thermal modes; c) one single-photon emitter and three thermal modes; d) one single-photon emitter, one Poissonian mode and one thermal mode and e) one single-photon emitter, one Poissonian mode and two thermal modes. The reconstructed optical fields without the presence of single-photon emitters are f) two thermal modes in presence of a Poissonian source; g) three thermal modes in presence of a Poissonian source; h) three thermal modes and i) four thermal modes.\\
It is clear from the plots and from the fidelities reported in Table I that our mode reconstruction method exploiting both $g^{(K)}$ and $\theta^{(K)}$ clearly outperforms the method using only the $g^{(K)}$ functions, as it correctly recognizes and reconstructs the type and number of light modes composing the optical fields under measurement and subsequent reconstruction.

\end{document}